\documentclass[12pt,a4paper]{article}

\usepackage{graphicx}
\usepackage{amssymb,amsmath}
\usepackage[cp1251]{inputenc}  %% 1
\usepackage[T2A]{fontenc}
\usepackage[russian, english]{babel}
\usepackage[unicode]{hyperref}
\usepackage{indentfirst}

\title{\bf On the neutron-antineutron oscillations}
\author{L.~B.~Okun\\
ITEP, Moscow, Russia}
\begin{document}
\maketitle
\begin{abstract}
This is a draft of my brief note on the early history of $n\bar{n}$ oscillations written  for the Project X at the request of Chris Quigg in March 2013
\end{abstract}

%Dear Chris,

%Thank you very much for your email of Feb 8, 2013 requesting one-page early history of n-nbar oscillations.  

The neutron-antineutron oscillations are of great interest because their observation would allow to test very accurately  the most fundamental CPT-symmetry, where C is charge  conjugation, P is inversion of space and T is reversal of time. The
CPT-symmetry is intimately connected with the name and life of Wolfgang Pauli, starting with his famous exclusion principle
\cite{pau1} and finishing with CPT-teorem \cite{pau2}.

The three discrete symmetries C, P, T were established soon after discovery of Quantum Mechanics.

C-symmetry or in other words symmetry between particles and corresponding antiparticles  was introduced in 1932 when the existence of positron (the antiparticle of electron) was predicted by Paul Dirac and discovered by Carl Anderson. 
Particles which are identical with their antiparticles are called
genuinely neutral. Such is the photon --- the particle of light; its C-parity is negative. A very important step forward was made by Murray Gell-Mann and Abraham Pais \cite{gmp} who introduced in the framework of C-symmetric theory the superpositions with positive and negative C-parity:
\begin{equation}
K_1=(K+\bar{K})|/\sqrt{2}
\end{equation}
\begin{equation} 
K_2=(K-\bar{K})/i\sqrt{2}
\end{equation}
As was shown by Abraham Pais and Oreste Piccioni \cite{pp} the propagation of $K_1$ and $K_2$ in the vacuum results in the vacuum oscillations between $K$ and $\bar{K}$

P-symmetry or in other words --- spatial parity --- the mirror symmetry between left-handed and right-handed objects and its violation was known to people long before the first scientific papers on the concepts of left and right appeared. In the XIX-th century the importance of left-right asymmetry (dissymmetry) for the processes of life became evident due to Luis Pasteur and others. With the advent of quantum mechanics in the 1920s  it was decided that biological dissymmetry is based on the  P-symmetry of the basic fundamental interactions. In the language using the concepts of Hamiltonian or Lagrangian that meant that scalar and pseudo-scalar terms in them cannot coexist. For almost 30 years it was believed that only scalar terms are present and no pseudoscalar terms could be added. This ended with the discovery of the discovery of the V-A weak current.

T-symmetry as well as P-symmetry was mainly formulated
in the framework of quantum mechanics around 1930 by Eugene Wigner. In the language using the concepts of Hamiltonian or Lagrangian the time-reversal invariance meant that the coupling constants of scalar and pseudocalar terms must be real. 

During 1956 Rochester conference Martin Block has asked in a private discussion with Richard Feynman his famous question: "Was Parity Conservation Proven?". Soon afterwards a negative reply appeared in a paper by Tsun-Dao Lee and Chen-Ning Yang who proposed \cite{ly} a list of experiments in which parity violation in a number of weak interaction processes was suggested. Very soon the very large P and C violating effects were confirmed by experiments and the 1957 Nobel Prize for this discovery was given to Lee and Yang. In their Nobel Prize talks \cite{y} \cite{l} they referred to the articles by Lee, Oehme, Yang \cite{loy} and Ioffe, Okun, Rudik \cite{ior} who indicated that violation of P was impossible without violation of C. The latter authors remarked also that the decay $K_2\rightarrow3\pi$ would mean that CP or T is conserved. Then Lev Landau postulated \cite{ll} that CP conservation is the most fundamental symmetry. 
But in 1964 the CP-violating decay $K_L\rightarrow\pi^+\pi^-$ was discovered \cite{turlay}. In an attempt to rescue the
arguments of Landau in favor of CP symmetry Kobzarev, Okun, Pomeranchuk introduced the concept of the mirror world \cite{kop}. The further evolution of this concept see in the review \cite{50y}. Unlike the mirror particles considered before \cite{ly} the mirror world of \cite{kop} is a dark matter world.

Lev Okun and Bruno Pontecorvo \cite{op} noticed that frequency of $K\bar{K}$ oscillations is very sensitive to the strength of interaction with $\Delta S=2$ and thus places the best upper limit on its admixture

Considerations of baryon number violation were triggered by the article by Andrey Sakharov on violation of CP and baryon asymmetry of universe \cite{ads}. It was followed by the remark by Vadim Kuzmin \cite{vak} that baryon number violation could lead to $n\bar n$ oscillations and by conclusion by Yuri Abov et al \cite{ado} that observation of these oscillations would represent the most accurate test of CPT-symmetry.See the reviews \cite{o1},\cite{o2},\cite{k}.

A lecture on connection between spin and statistics see \cite{o3}.

I am grateful to Chris Quigg for requesting me to write this brief text for the Project X and to Yuri Kamyshkov for refreshing some referenes and the following suggestion: 

"`In theoretical papers there are many models that discuss n-nbar, like
GUT, SUSY, B-L violation, seasaw models, majorana neutron, extra-dimension
models, models predicting possible new scalars at LHC,  models with
low-scale baryogenesys and some more. They are also part of more recent
history of n-nbar that has been developed after Sakharov  and Kuzmin. I guess
that you possibly see your paper as an early history part of n-nbar story
and assume that Chris Quigg and his "n-nbar white paper" team will
review these models. However, if you will decide to post this paper to
arXiv as a separate paper you might like to mention more recent theoretical
models that you are not covering and that were partialy reviewed by Rabi Mohapatra
in \url{http://arxiv.org/pdf/0902.0834.pdf}"'

\end{document}